\documentclass[12pt]{article}

\usepackage{latexsym}
\usepackage{graphics}
\usepackage{epsfig}
\usepackage[mathscr]{eucal}
\usepackage{amscd}
\usepackage{amssymb}
\usepackage{amsmath}
 
\newcommand{\fB}{\mathfrak A}
\newcommand{\cB}{{\mathfrak B}} 

\newcommand{\cP}{{\mathcal P}}
\newcommand{\sq}{\sqcup}
 
\newcommand{\bZ}{{\mathbb Z}} 
\newcommand{\cA}{{\mathcal A}}
\newcommand{\cC}{\mathfrak C} 
\newcommand{\defn}{\equiv}  
\newcommand{\cM}{\mathcal M}
\newcommand{\card}{\rm{card}}
 
\newcommand{\tG}{\widetilde \Gamma}
\newcommand{\al}{\alpha} 

\newcommand{\supp}{\rm {supp}} 

\newtheorem{conjecture}{Conjecture}

\newtheorem{lemma}{Lemma}
\newtheorem{definition}{Definition}
\newcommand{\bproof}{\setlength{\parindent}{0mm}{\bf Proof{~~}}}
\newcommand{\eproof}{\hfill $\Box$\setlength{\parindent}{5mm}}
\newcommand\beq{\begin{equation}}
\newcommand\eeq{\end{equation}}
\newcommand\bea{\begin{eqnarray}}
\newcommand\eea {\end{eqnarray}}

\newcommand{\upA}{\uparrow \! \! A}
\newcommand{\downA}{\downarrow \!\! A}
\newcommand{\nin}{\not\in}
\newcommand{\tgamma}{\widetilde \gamma}
\newcommand{\tP}{\widetilde P}
 
\def\nn{\nonumber} 
 
\title{Quantum Covers in Quantum Measure Theory} 
\author{Sumati Surya \& Petros Wallden \\
Raman Research Institute, Bangalore, India }
\begin{document}
\baselineskip 20pt
\maketitle

\begin{abstract}

In standard measure theory the measure on the base set $\Omega$ is
normalised to one, which encodes the statement that ``$\Omega$
happens''. Moreover, the rules imply that the measure of any subset
$A$ of $\Omega$ is strictly positive if and only if $A$ cannot be
covered by a collection of subsets of zero measure.  In quantum
measure theory on the other hand, simple examples suffice to
demonstrate that this is no longer true. We propose an appropriate
generalisation of a cover to quantum measure theory, the {\sl quantum
cover}, which in addition to being a cover of $A$, satisfies the
property that if every one of its elements has zero quantum measure,
then so does $A$. We show that a large class of inextendible
antichains in the associated powerset lattice provide quantum covers
for $\Omega$, for a quantum measure that derives from a strongly
positive decoherence functional.  Quantum covers, moreover, give us a
new perspective on the Peres-Kochen-Specker theorem and its role in
the anhomomorphic logic approach to quantum interpretation. 

\end{abstract}

\section{Introduction} 

One of the goals of the decoherent histories approach to quantum
theory is to formulate precisely how classicality emerges from quantum
theory \cite{gellmannhartle}. A partition $\Omega_c$ of the space of
histories $\Omega$ is considered to be classical if the decoherence
functional $D(A,B)$ between any two distinct elements $A,B \in
\Omega_c$ is weakly zero. This allows one to confer to $D(A,A)$ the
status of a classical probability, or measure, obeying the standard
probability sum rules. To complete this picture, the decoherence
functional on the full histories space $D(\Omega,\Omega)$ must be
strictly positive, and thus normalisable to 1. From a realist's
perspective, this gives us the coarsest possible sense in which we can
say that ``something happens''.

The {\sl quantum measure} is a natural generalisation of the classical
measure and is defined for {\it any} subset $A \subseteq \Omega$ to be
$|A|=D(A,A) \geq 0$,\footnote{This inequality is often referred to as
the condition of ``positivity'' which we will henceforth assume for a
quantum measure.} with the same requirement $D(\Omega,\Omega)>0$
\cite{qmeasure}. This measure does not satisfy the Kolmogorov sum rule
since it can involve a non-vanishing interference term
\begin{equation} 
I_2(A,B)=|A\sqcup B| - |A|-|B| 
\end{equation} 
for a pair of disjoint sets $A,B \subset \Omega$, where $\sqcup$
indicates disjoint union. Instead, as shown in \cite{qmeasure}, the
quantum measure satisfies the quantum sum rule
\begin{equation}
I_3(A,B,C)=|A\sq B \sq C|-|A\sq B| -|A \sq C| - |B\sq C| +
|A|+|B|+C|=0.     
\end{equation} 
Thus, while the classical measure satisfies the null test for the two
slit experiment with $I_2(A,B)=0$, the quantum measure in standard
quantum theory satisfies the null test for the three slit, with
$I_3(A,B,C)=0$. The generalisation to the $n$-level interference term
$I_n(A_1, \ldots , A_n)$ on $n$-tuples is obvious, and standard
quantum theory satisfies a null test for all $n$-slit experiments,
with $n>2$.  It was shown in \cite{qmeasure} that if the $n$th-level
interference vanishes on all $n$-tuples, then so do all higher level
interferences.  This implies a hierarchy of generalised quantum
measures, with classical theory being level 1, and standard quantum
theory being level 2.  We will refer to as ``standard'' a quantum
measure of level 2.

In classical measure theory the condition $|\Omega|_c>0$\footnote{ We
will use $|.|_c$ to denote the classical measure.} puts the following,
somewhat trivial restriction on the choice of measure for the subsets
of $\Omega$. Namely, if $\Omega$ admits a covering of zero measure
sets, then $|\Omega|_c=0$, and hence such {\sl zero covers} are
disallowed. On the other hand, zero covers are allowed in quantum
measure theory. As an example, consider the 3 slit experiment and
restrict to three paths $A,B$ and $C$, one through each slit, all of
which arrive at the same non-dark spot on the screen. It is possible
for $|A\sqcup B|=0$ and $|B\sqcup C|=0$, but $|A\sqcup C|\neq 0$, so
that $|\Omega| >0$. Thus, while the pair of subsets $A\sqcup B$ and $
B \sqcup C$ suffices to cover $\Omega$, this does not imply that
$|\Omega|=0$. In this sense, the pair $\{ A\sqcup B, B\sqcup C\}$
doesn't ``cover'' $\Omega$ sufficiently, when using the quantum
measure. We use this idea to define the  {\sl quantum cover} of $\Omega$ as:

\begin{definition} For a quantum measure $|.|$, $\{O_i \}$ is said to
  be a {\sl quantum cover} of $\Omega$ if $\Omega= \bigcup_i O_i$ and
  $|O_i|=0$ for all $i$ implies that $|\Omega|=0$.
\end{definition} 

Does a non-trivial quantum cover of $\Omega$ always exist? We begin in
section (\ref{main}) by showing that it does.  The quantum cover finds
its rightful place in the powerset lattice $\cB$ associated with the
powerset $2^{|\Omega|}$, where set inclusion provides the order
relation, and the minimal element is the empty set and the maximal
element is $\Omega$. We prove existence by showing that any $k$-level
inextendible antichain in $\cB$ satisfies the criterion for a quantum
cover, for {\it any} quantum measure, when $k \geq 2$.  Motivated by
this, we examine more general classes of inextendible antichains in
$\cB$ and show that they also satisfy this criterion, when the quantum
measure is derived from a strongly positive decoherence functional
\cite{qrw}\footnote{The assumption of strong positivity is natural --
all known physical systems satisfy it.}. This is summarised as
\begin{lemma}\label{result} Let $\cA$ be an
  intextendible antichain in the powerset lattice $\cB$ over a finite
  histories space $\Omega$, which belongs to the class $\cC$. If each
  of the elements in $\cA$ has zero quantum measure, then
  $|\Omega|=0$, when $|.|$ is assumed to come from a strongly positive
  decoherence functional \footnote{And possibly, more general quantum
  measures, which are only required to satisfy conditions
  (\ref{first}) and (\ref{second}) (see Section \ref{main}). }. Thus
  $\cA$ is a quantum cover of $\Omega$.
\end{lemma} 
Inextendible antichains in $\cB$ do not possess a universal
characterisation, apart from satisfying somewhat weak inequalities
\cite{sperner}, and hence it is unclear how to proceed most
generally. However, the class $\cC$ examined so far is general enough
for us to conjecture a deeper connection between inextendible
antichains and quantum coverings: 
\begin{conjecture} \label{conjecture}
For a quantum measure that is obtained from a strongly positive
decoherence functional\footnote{Or more generally one which satisfies
(\ref{first}) and (\ref{second}).}, every inextendible antichain in
$\cB$ is a quantum cover of $\Omega$.
\end{conjecture}

We end this section with a brief diversion to the preclusion-based
anhomomorphic logic interpretation of quantum theory.  In particular,
we point out the existence of a special inextendible antichain in the
powerset lattice which spurred our search for a quantum cover.
 
One of the motivations for constructing a quantum measure theory is to
be able to use it in a manner analogous to the classical measure,
i.e., to say something more about physical reality than the standard
Copenhagen interpretation allows. For example, one would like to be
able to interpret a set of histories of zero measure to unambiguously
imply that they do not occur.  However, contrary to classical
intuition, such sets can contain subsets of non-vanishing measure. An
example of this is the two slit experiment, for which pairs of
destructively interfering histories have zero measure, although each
individual history has non-zero measure.

The space of subsets of the histories space $\Omega$ or
the power set $2^{|\Omega|}$ forms a unital Boolean algebra $\fB$,
with addition $A+B$ defined as symmetric difference, multiplication
$AB$ as set intersection, and with $\Omega$ being the unit element.
Classical logic involves a homomorphism or {\sl coevent} $\Phi_c$ from
$\fB$ to ${\bZ}_2$, the set of truth values, with
\begin{equation} 
\Phi_c(A+B)= \Phi_c(A) + \Phi_c(B), \quad
\Phi_c(AB)=\Phi_c(A)\Phi_c(B), \quad 
\Phi_c(\Omega) =1.  
\end{equation} 
An actualisation or reality is a {\sl primitive} coevent, $\Phi_c^p$,
i.e., one whose support $\supp(\phi_c^p)$ is a single fine grained
history. Any primitive coevent $\phi_c^p$ for a classical system is
also ``preclusive'': for all sets $P \in \fB$ of zero measure
$|P|_c=0$,
\begin{eqnarray} 
\Phi_c^p(P) &= & 0, \label{cpreclusion} \\ {\rm{supp}} (\Phi_c^p)
&\nsubseteq& P. \label{pmodusponens}
\end{eqnarray} 
In other words, sets of zero measure cannot be realised.

The anhomomorphic logic proposal \cite{alogic} generalises classical logic
to coevents $ \Phi: \fB \rightarrow {\bZ}_2$ which are {\it not}
homomorphisms (hence the term ``an-homomorphic''), but are {\sl
preclusive} in that they satisfy Eqn (\ref{cpreclusion}).  In
addition, $\Phi$ could at best retain a part, but not all, of the
Boolean structure of $\fB$. In the {\sl multiplicative scheme} one
retains
\begin{equation} 
\Phi(AB)=\Phi(A)\Phi(B) 
\end{equation} 
so that a preclusive coevent also  satisfies  the ``modus ponens'' of
classical logic 
\begin{equation} 
\Phi(A)=1  \Rightarrow \Phi(B)=1 \, \, \forall \, \, B \supset A.  
\end{equation}  
This means that not only is $\Phi(\Omega)=1$, but Eqn
(\ref{pmodusponens}) is also satisfied. Thus, a zero measure set
always maps to the zero element in $\bZ_2$, i.e., is
``false''. Moreover, every subset of this set also maps to the
zero element.  Thus, even though a set $P$ of zero quantum measure
can contain a subset $Q \subset P$ of non-zero quantum measure,
neither $P$ nor $Q$ can be realised. From the example of the double
slit, it means that for two destructively interfering paths $A$, $B$,
not only is $A+B$ false, but so are $A$ and $B$, individually, even
though their quantum measures are strictly non-zero. Again, in analogy
with the classical system, quantum reality is represented by a {\sl
primitive preclusive coevent}(PPC), i.e., preclusive coevents with the
smallest possible support. The supports of PPCs are typically not fine
grained histories and hence quantum reality can manifest itself at
best as a collection of fine grained histories. As shown in
\cite{dgt} this scheme passes the stringent test of the
Kochen-Specker theorem, and hence has emerged as a promising candidate
for a realist interpretation of quantum theory.

Our proposal for a quantum cover comes from the observation that the
set of supports of the set of PPC's forms an antichain $\cA$ in
$\cB$. The up set $\upA \defn \{b \supseteq a| a \in \cA \} $ contains
no zero measure sets, and hence all zero measure sets must lie either
to the past of $\cA$, or be incomparable to it.  In particular, the
set of maximal elements $\cM' $ of $\cB' \equiv \cB \backslash \upA $
is an antichain in $\cB$ containing only zero measure sets: any $m$
that is not of zero measure either lies in $\uparrow \cA$ or is
contained in a zero measure set. $\cM'$ is thus made up of the largest
possible zero measure sets.

The maximal elements of $\cB \backslash (\upA \backslash \cA)$ forms
an antichain $\cA'$ in $\cB$, with $\cA' = \cA \sqcup \cM$ and  $\cM
\subseteq \cM'$. Any element of $\cM'$ which is not in $\cM$ must lie
in the down set $\downA$ (defined dually to $\upA$). Moreover, since
every zero measure set either lies in $\cM'$ or is contained in an
element of $\cM'$, this means that every zero measure set lies in
$\downA'$. If a set is not precluded, then it either lies in $\upA$ or
is contained in an element of $\cM'$ and hence contained in an element
of  $\cM$ or $\cA$.

Thus, $\cA'$ is an inextendible antichain in $\cB$, with a portion of
the elements reserved for what doesn't happen, and the remaining for
what can; it is crucial that that the set of possible realisations
(i.e., the supports of PPC's) are {\it not} of zero measure, i.e. that
$\cA'$ is not a zero cover. A natural question that arises from this
analysis is whether, for a physically realisable quantum system $\cB$
admits an inextendible antichain containing only elements of
zero measure. We make considerable progress in proving otherwise.

\section{Constructing Quantum Covers} 
\label{main} 
Let $A_1, A_2 \ldots A_n$ represent the fine grained histories in
$\Omega$. The associated powerset lattice of $\Omega$ is obtained by
taking the empty set as the bottom element and $\Omega$ as the top
element, with the order corresponding to set inclusion. Thus, $X \prec
Y $ in $\cB$ iff $ X \subset Y$ in $\Omega$. Cardinality of a set
defines a ``level'' in $\cB$ -- for example, $(A_1\sqcup A_2)$ is a
level 2 element in $\cB$.

As mentioned in the introduction, a $k$-level quantum measure  is defined
as the smallest  $k$ for which the $k+1$ way interference 
\begin{eqnarray} 
I_{k+1}(A_1, \ldots A_{k+1}) & = &  |A_1\sqcup  \ldots \sqcup A_{k+1}|
\nn \\ 
&&   -\sum_{k-\mathrm{subsets} } |k-\mathrm{subset}| +
\sum_{(k-1)-\mathrm{subsets}} |(k-1)-\mathrm{subset}| \ldots \nn \\ 
&& + \, (-1)^{k+2} [|A_1| + |A_2| \ldots +|A_{k+1}|],     
\end{eqnarray} 
is zero, where by a $j$-subset we mean a $j$-element subset of $ \{
A_1, \ldots A_{k+1}\}$, $j=1,\ldots k+1$, and
$\sum_{j-\mathrm{subset}}$ is a sum over all possible $j$-subsets.  If
so, then all higher interference terms also vanish
\cite{qmeasure,rob}, and one obtains, for $n>k$ the identity
\begin{eqnarray} 
 |A_1\sqcup  \ldots \sqcup A_{n}| & =& \sum_{(n-1)-\mathrm{subsets}} |(n-1)-\mathrm{subsets}| - 
\sum_{(n-2)-\mathrm{subsets}} |(n-2)-\mathrm{subsets}| \nn \\ 
&& \ldots + (-1)^{n} [|A_1| + |A_2| \ldots +|A_{n}|].      
\end{eqnarray} 
Using this identity recursively, one can show inductively
that\footnote{Proof in the Appendix} for a $2$-level quantum
measure
\begin{equation} \label{identity} 
 |A_1\sqcup  \ldots \sqcup A_{n}| = (2-n)\sum_{i=1}^n |A_i|
  +\frac{1}{2} 
  \sum_{i,j=1}^n |A_i\sqcup A_j|.   
\end{equation} 
In order to avoid confusion with the use of the word ``level'' in
$\cB$, we fix the quantum measure to be 2-level once and for
all\footnote{A similar type of identity may be obtained for higher
level quantum measures.}.  We will make crucial use of the following
identities
\begin{equation} \label{first} 
|A\sqcup B| = 0 \Rightarrow |A|=|B|
\end{equation} 
and 
\begin{equation} \label{second} 
|A|=0 \Rightarrow |A\sqcup B|=|B|.  
\end{equation} 
which follow from the strong positivity of the  decoherence
functional, as shown in the Appendix.

\subsection{Proof of existence}

We show that the simplest example of an inextendible antichain $\cA$
in $\cB$, namely the set of all level $k$ elements, with $0<k<n$
provides a quantum cover for $\Omega$, for any choice of quantum
measure. 

Let each of the elements in $\cA$ be of zero measure. If $k=1$, then
since each $|A_i|=0$, by Eqn (\ref{second}) $|\Omega|=|A_1 \sqcup A_2
\ldots \sqcup A_n|=0$.  Let $k \geq 2$. Using Eqn (\ref{identity}) for
a level $k$ element $(A_{i_1}, A_{i_2} \ldots A_{i_k})$ of $\cB$,
\begin{equation}
|A_{i_1}\sqcup  A_{i_2} \ldots \sqcup A_{i_k}|=(2-k)(|A_{i_1}| +
 \ldots |A_{i_k}|) + (|A_{i_1}\sqcup A_{i_2}| + \ldots +
 |A_{i_k-1}\sqcup A_{i_k}|)=0.      
\end{equation} 
Adding  up all such $\binom{n}{k}$ terms, we obtain 
\begin{eqnarray} 
\binom{n-1}{k-1} (2-k) \sum_{i=1}^n |A_i| + \binom{n-2}{ k-2}
\sum_{i,j=1, i< j}^n |A_i \sqcup A_j|& = & 0 \\ 
\Rightarrow   \sum_{i,j=1, i< j }^n |A_i\sqcup A_j| = 
\frac{(n-1)(k-2)}{(k-1)} \sum_{i=1}^n |A_i|. &&     
\end{eqnarray} 
Substituting in the  expression Eqn (\ref{identity}) for $|\Omega|=
|A_1 \sqcup A_2 \ldots \sqcup A_n|$ 
\begin{equation} 
|\Omega|= \biggl( (2-n) + \frac{(n-1)(k-2)}{(k-1)}
 \biggr) \sum_{i=1}^n |A_i| \leq 0 \Rightarrow |\Omega|=0,    
\end{equation} 
since the coefficient simplifies to  $\frac{(k-n)}{(k-1)} < 0$ for
$2\leq k<n$. This proves 
\begin{lemma} 
The level $k\geq 2 $-antichain in $\cB$ is a quantum cover of $\Omega$. 
\label{lemone}
\end{lemma} 
Note that the above Lemma doesn't require strong positivity. However,
to prove that the $k=1$ antichain is a quantum cover, we did need 
strong positivity.  We now generalise this result to a large class of
inextendible antichains obtained by systematically ``whittling'' away
at the level $k$ antichain.

\subsection{Generalisation} 

The powerset lattice has a great deal of structure: every $k$-level
element has exactly $k$-links to level $k-1$ and exactly $n-k$ links to level
$k+1$. Thus, any antichain obtained by removing $m<k+1$ elements in
the $k$-level has a unique completion to the $k$-level antichain if $n
\geq 2k$ and similarly if $m< n-k+1$ for $n\leq 2k$.

We define the past and future shadows of an element $a \in \cB$ on the
$k$-level antichain to be
\begin{eqnarray} 
S_k(a)\equiv \{ b \subset a | \card(b)=k\} &\qquad& \card(a)>k \nn \\
S_k(a)\equiv \{ b \supset a | \card(b)=k\} &\qquad& \card(a)<k,  
\end{eqnarray} 
where $\card(a)$ denotes the cardinality of the set $a$. The strategy
we employ is to whittle away at the $k$-level antichain by (a) picking
a set of $m$ mutually unrelated elements $\Lambda'(\neq k) \equiv \{
a_s\}$ from levels other than $k$ (b) adding in elements in the
$k$-level antichain $\Lambda(k)$ which do not lie in the shadows of
the $a_s$ (c) finding an inextendible extension of the resulting
antichain by adding more elements to $\Lambda'(\neq k)$ to get
$\Lambda(\neq k)$, without changing $\Lambda(k)$, or equivalently,
without changing the set of elements on the $k$-level antichain which
lie inside the shadows of the elements of $\Lambda'(\neq k)$. The
resulting antichain is then $\cA= \Lambda(k) \sqcup \Lambda(\neq k)$.

To illustrate, let us start with a single element, $\Lambda( k+1)=
\{a_1\}$ in level $k+1$. Wlog, let $a_1=(A_1 \ldots A_{k+1})$. Then
$\Lambda(k)=\{ (A_{\alpha_1}, \ldots
A_{\alpha_{k-r}},A_{\beta_1},\ldots, A_{\beta_r})\} $ is the set of
$\binom{n}{k} - (k+1)$ elements of the $k$-level antichain which do
not lie in the shadow of $a_1$, where $\alpha_i \in [1, \ldots k+1] $,
$\beta_j \in [k+2, \ldots n]$, and $r\in [1, \ldots r_0]$ with $r_0=
\min(k,n-k-1)$. $\cA$ is clearly inextendible, since every $k$ level
element either lies in $\cA$ or is contained in an element of $\cA$,
and every $k+1$ level element is either in $\cA$ or contains an
element of $\cA$.

For  the $k+1$ elements in $S_k(a_1)$, using  Eqn (\ref{first})
\begin{eqnarray} \label{setone} 
|A_{\alpha_1} \sqcup A_{\alpha_2 } \ldots \sqcup A_{\alpha_{k}}| & = &
|A_{\alpha_{k+1}}| \nn \\ 
\Rightarrow (2-k)\sum_{i=1}^{k} |A_{\alpha_i}| + \frac{1}{2} 
\sum_{i,j =1}^{k} |A_{\alpha_i}\sqcup A_{\alpha_j}| &=
&|A_{\alpha_{k+1}}|
\end{eqnarray} 
For the remaining $\binom{n}{k}-(k+1)$ elements in $\Lambda(k)$, 
\begin{eqnarray} 
\label{settwo} 
(2-k) \bigl( \sum_{i=1}^{{k-r}} |A_{\alpha_i}| +
    \sum_{j=1}^{r}|A_{\beta_j}|\bigr) +  
    \frac{1}{2}\sum_{i,j=1}^{k-r}|A_{\alpha_i} \sqcup A_{\alpha_j}|+ && \nn
    \\ 
    \sum_{i1}^{k-r}\sum_{j=1}^{r} |A_{\alpha_i} \sqcup A_{\beta_j}| +  
    \frac{1}{2}\sum_{i,j=1}^{r}|A_{\beta_i} \sqcup A_{\beta_j}| & = &  0,  
\end{eqnarray} 
where $\alpha_i \in [1, \ldots k+1]$ and $\beta_j \in [1, \ldots, r]
$. 
Adding all the $\binom{n}{k}$ equations (\ref{setone}) and (\ref{settwo}) 
\begin{equation} 
(2-k) \binom{n-1}{k-1} \sum_{i=1}^n |A_i| + \binom{n-2}{k-2}
   \frac{1}{2}\sum_{i,j=1}^n |A_i\sqcup A_j| = \sum_{i=1}^{k+1} |A_i|  
\end{equation} 
Solving for the term $ \frac{1}{2} \sum_{i,j=1}^n |A_i\sqcup A_j|$ and
inserting into the rhs of Eqn (\ref{identity}) we find 
\begin{eqnarray} 
|\Omega| & =&  \biggl[(2-n) +\frac{(k-2)(n-1)}{(k-1)} +\frac{1}{
    \binom{n-2}{k-2}}\biggr]\sum_{i}^{k+1}|A_{\alpha_i}| \nn \\  
&& + \biggl[(2-n) +
 \frac{(k-2)(n-1)}{(k-1)}\biggr] \sum_{i=k+2}^n |A_i| \leq 0,  
\end{eqnarray} 
which means that $|\Omega|=0$.

Alternatively, we can choose $\Lambda(k-1)= \{ a_1\}$ to be an element
in the $k-1$-level, i.e.  $|A_1 \sqcup A_2 \ldots \sqcup
A_{k-1}|=0$. $S_k(a_1)$ is now the set of $k$-level elements that
contain $a_1$ as a subset, i.e. all those of the form $(A_1, A_2
\ldots, A_{k-1},A_i)$, with $i \in [ k, \ldots n]$. Thus 
\begin{equation} \label{kminus} 
|A_1 \sqcup A_2 \ldots \sqcup A_{k-1} \sqcup A_i|=|A_i| \, \forall
 \,\, i \in [k, \ldots, n].  
\end{equation} 
The remaining elements in $\Lambda(k)$ have measure zero, so that 
again,  adding up all the measures of the $k$-level elements we get  
\begin{equation}
 (2-k) \binom{n-1}{k-1} \sum_{i=1}^n |A_i| + \binom{n-2}{k-2}
   \frac{1}{2}\sum_{i,j=1}^n |A_i\sqcup A_j| = \sum_{i=k}^{n} |A_i| 
\end{equation} 
which implies 
\begin{eqnarray} 
|\Omega| & =& + \biggl[(2-n) + \frac{(k-2)(n-1)}{(k-1)}\biggr]
 \sum_{i=1}^{k-1} |A_i| \nn \\ && \biggl[(2-n)
 +\frac{(k-2)(n-1)}{(k-1)} +\frac{1}{\binom{n-2}{k-2}}\biggr]
 \sum_{i=k}^n |A_{i}| \leq 0, 
\end{eqnarray} 
or  $|\Omega|=0$ as before.

The strategy for this line of reasoning should by now be clear.
Instead of the $k$-level equations,  one could ``project'' onto any
level $2 < l < n$. Summing over the measure of all the level $l$
elements gives the now recognisable term 
\begin{equation}\label{klevelsum} 
(2-l) \binom{n-1}{l-1} \sum_{i=1}^n |A_i| +
    \binom{n-2}{l-2}\frac{1}{2} \sum_{i,j=1}^n |A_i\sqcup A_j|,   
\end{equation} 
which we try to evaluate using the details of the inextendible
antichain $\cA$: the measure of each $l$-level element $a_l$, can be
determined by knowing how it relates to an element of $\cA$. If $a_l
\geq a \in \cA$, then strong positivity implies that
$|a_l|=|a_l\backslash a|$ as in Eqn (\ref{setone}) and if $a_l \leq a
\in \cA$, then $|a_l|=|a\backslash a_l|$, as in Eqn 
(\ref{kminus}). This should help us simplify the expression for
$|\Omega|$ sufficiently to prove the result.  This procedure clearly
depends crucially on the details of $\cA$.

We now focus on particular classes of generalisations for which
$\Lambda(k)$ is ``sufficiently'' populated in the following sense.  It
is useful to define an {\sl index set} $\Gamma_s$ associated with any
$l$-level element $a_s \in \cB$ as the set of labels $\{
\alpha^{(s)}_j \}$, $j=1, \ldots l$, where $a_s=A_{\alpha_1} \sqcup
A_{\alpha_2} \ldots A_{\alpha_l}$. We will reserve the symbol $\Gamma$
for the complete set of labels $\{1, \ldots, n\} $.  Define $\tG
\equiv \Gamma\backslash \bigcup_{s} \Gamma_s$, where $a_s \in
\Lambda(\neq k)$ and $p\equiv |\tG|$. The requirement that
$\Lambda(k)$ be sufficiently populated is the requirement that there
be a lower bound on $p$. We will consider the three cases separately,
$\Lambda(>k)$, $\Lambda(<k)$ and the mixed case $ \Lambda(>k) \sqcup
\Lambda(<k)$.

\subsubsection{$\cA=\Lambda(k)\sqcup \Lambda(>k)$} \label{greaterk}

Let $\Lambda(>k)\equiv \{a_s \} $, $s \in [1 , \ldots , m]$ and $p
\geq 1$. This means that there is at least one fine grained history
$A_1$ which is contained only in some of the elements of $\Lambda(k)$
but not in any element of $\Lambda(>k)$.  The set of all $k$-level
elements containing $A_1$ has zero measure, i.e. $|A_{\alpha_1}\sqcup \ldots
  A_{\alpha_{k-1}}\sqcup A_1| =0$,  $\forall \, \, {\alpha_i} \neq
  1$. This implies that 
\begin{equation} 
  |A_{\alpha_1}\sqcup \ldots A_{\alpha_{k}}\sqcup A_1| = 
  |A_{\alpha_1}| = \ldots =|A_{\alpha_k}| \label{pgeqonedoublet}
\end{equation} 
which means that all $|A_{\alpha_i}|$'s are  equal for all $\alpha_i
\neq 1 $. Call this $|A_\alpha|$. Moreover, 
\begin{equation} 
 |A_{\alpha_1}\sqcup \ldots A_{\alpha_{k+1}}\sqcup A_1|=|A_{\alpha_1}
  \sqcup A_{\alpha_2}|= \ldots = |A_{\al_{k}}\sqcup A_{\al_{k+1}}|
\end{equation} 
from which we deduce that all $|A_{\al_i} \sqcup A_{\al_j}|$'s are equal
for $\al_i, \al_j \neq 1 $. Call this $|A_\alpha \sqcup
A_{\alpha'}|$. Using (\ref{identity}), the $k$-level zero measure sets
give 
\begin{equation} 
(2-k)|A_1|+ (2-k)(k-1) |A_{\alpha}|+ \sum_{i=1}^{k-1} |A_1 \sqcup A_{\al_i}| +
  \frac{(k-1)(k-2)}{2}|A_\alpha \sqcup A_{\alpha'}| =0.       
\end{equation} 
Summing over all $\binom{n-1}{k-1}$ of these 
\begin{equation} 
 (2-k)|A_1|+ (2-k)(k-1) |A_{\alpha}|+
\frac{k-1}{n-1} \sum_{i=2}^{n} |A_1 \sqcup A_i| 
 +\frac{(k-1)(k-2)}{2}|A_\alpha \sqcup A_{\alpha'}|   =  0.  
\end{equation} 
Moreover, for any $a_s \in \Lambda(>k)$, $\{ \alpha_i \}=\Gamma_s $ 
\begin{equation}
|A_{\alpha_1} \sqcup \ldots \sqcup A_{\alpha_s}|=0  \Rightarrow
 \frac{1}{2}|A_\alpha \sqcup A_{\alpha'}| = \frac{s-2}{s-1}|A_\alpha|.   
\end{equation} 
Inserting into (\ref{identity}) we get 
\begin{equation}
|\Omega| = (k-n) \biggl( \frac{1}{k-1}|A_1| + \frac{n-1}{s-1}|A_\alpha|
 \biggr) \leq 0. 
\end{equation} 

\subsubsection{$\cA=\Lambda(k)\sqcup \Lambda(<k)$}  \label{lessk} 

Let $s_0$ be the lowest level in $\Lambda(<k)$, and let $r= k -
s_0+1$. Let $p \geq r$, so that at least $r$ fine grained histories
$A_1, \ldots A_r$ are not contained in any of the elements of
$\Lambda(<k)$.  Define $P \equiv \{ 1, \ldots r\} \subset \Gamma$. 
For any set $Q \subset \Gamma\backslash P$, with $q\equiv |Q| = s_0 -1$,
$P\sqcup Q \not\supset \Gamma_s$ for any  $s$, by construction. Hence, it
is an index set of an element in $\Lambda(k)$  
\begin{equation} 
|A_1 \sqcup \ldots A_r \sqcup A_{\alpha_1} \sqcup \ldots
 A_{\alpha_q}|  =   0, \, \, \alpha_i \nin P  \label{kklevel}. 
\end{equation} 
Since  $n \geq k+1$ 
\begin{equation} 
|A_1\sqcup \ldots A_r\sqcup  A_{\alpha_1} \sqcup \ldots A_{\alpha_{q+1}}| = 
  |A_{\alpha_1}| = \ldots =|A_{\alpha_q+1}| 
\end{equation} 
which means that all $|A_{\alpha_i}|$'s are equal for $\alpha_i \nin
P$. Call this $|A_\alpha|$. If, in addition, $n \geq k+2$ then
\begin{equation} 
|A_{1}\sqcup \ldots A_r\sqcup  A_{\alpha_1} \sqcup \ldots A_{\alpha_{q+2}}| = 
  |A_{\alpha_1}\sqcup A_{\alpha_2}|
  = \ldots
  =|A_{\alpha_{q+1}} \sqcup A_{\alpha_{q+2}}|\label{pairwise}, 
\end{equation} 
which means that all $|A_{\al_i} \sqcup A_{\al_j}|$'s are equal for
$\al_i, \al_j \nin  P$. We call this $|A_\alpha \sqcup
A_\alpha'|$ as before. Using (\ref{identity}), (\ref{kklevel}) reduces
to
\begin{eqnarray} 
(2-k) \sum_{i=1}^r |A_i|+ (2-k)q
  |A_{\alpha}|+ \frac{1}{2}\sum_{i,j=1}^r
  |A_i\sqcup A_j | + && \nonumber \\ 
\sum_{i=1}^r \sum_{j=1}^{q} 
|A_i \sqcup A_{\alpha_j}| + 
  \frac{q(q-1)}{2}|A_\alpha \sqcup
  A_{\alpha'}| &= & 0.       
\end{eqnarray} 
Summing over all $\binom{n-r}{q}$ of these, gives us 
\begin{eqnarray} 
\binom{n-r}{q} \biggl( (2-k) \sum_{i=1}^r |A_i|+ (2-k)q
|A_{\alpha}|+ \frac{1}{2}
\sum_{i=1}^r \sum_{j=1}^r
  |A_i\sqcup A_j | + && \nonumber \\ 
\frac{q(q-1)}{2}|A_\alpha \sqcup A_{\alpha'}| \biggr) + \binom{n-r-1}{q-1}
\sum_{i=r+1}^n \sum_{i=1}^{r} |A_i
\sqcup A_j|&= & 0. \label{sumover}  
\end{eqnarray} 
Moreover, for any $a_s \in \Lambda(<k)$, $\{\alpha_i\} \in \Gamma_s$, 
\begin{equation}
|A_{\alpha_1} \sqcup \ldots \sqcup A_{\alpha_s}|=0  \Rightarrow
 \frac{1}{2}|A_\alpha \sqcup A_{\alpha'}| = \frac{s-2}{s-1}|A_\alpha|. \label{as}  
\end{equation} 
Inserting into (\ref{identity}) we get 
\begin{eqnarray} 
|\Omega| & = &   \frac{k-n}{q}\biggl( (2-r) \sum_{i=1}^r|A_i| +
\frac{1}{2} \sum_{i=1}^r \sum_{j=1}^r |A_i\sqcup A_j | \biggr) 
+ \frac{(k-n)(n-r)}{s-1} |A_\alpha| \nn \\ 
&=&  (k-n)\biggl( \frac{1}{q} |A_1\sqcup \ldots \sqcup A_r| +
\frac{(n-r)}{s-1} |A_\alpha| \biggr)\nn \\ 
&\leq & 0. \label{zetas}   
\end{eqnarray} 
For $k=n-1$, we cannot use the simplification (\ref{pairwise}). Noting
that $\Omega=A_1\sqcup \ldots A_r \sqcup A_{r+1} \sqcup \ldots A_n$,
we see that $\Lambda(<k)$ consists of the single element
$a_s=A_{r+1} \sqcup \ldots A_n$ in level $s_0=n-r$. Replacing 
\begin{equation} 
\binom{n-r}{q}\frac{q(q-1)}{2}|A_\alpha\sqcup A_{\alpha'}| \rightarrow   
\binom{n-r-2}{q-2}\frac{1}{2} \sum_{i,j=r+1}^n |A_i \sqcup A_j|
\end{equation} 
in Eqn (\ref{sumover}) and replacing (\ref{as}), with
\begin{equation}
\frac{1}{2}\sum_{i=r+1}^n \sum_{j=r+1}^n |A_i \sqcup A_j| = 
(n-r)(n-r-2)|A_\alpha|,   
\end{equation} 
we recover Eqn (\ref{zetas}) with $s=s_0$,  thus proving our result for all $k$.

\subsubsection{$\cA=\Lambda(k)\sqcup
  \Lambda(<k) \sqcup \Lambda(>k)$} \label{mixedk} 

Again, define $r$ and $q$, $P$ and $Q$ via the lowest level $s_0$ in
$\Lambda(<k)$. Then $P\sqcup Q \not \supset \Gamma_s$ for any
$\Gamma_s$ coming from $\Lambda(<k)$ and also, since $P \cap
\Gamma_s=\emptyset$ for all $s$, $P\sqcup Q \not \subset \Gamma_s$ for
any $\Gamma_s$ coming from $\Lambda(>k)$. Hence the construction goes
through as in Section \ref{lessk}]. 

In particular, if we redefine $s_0$ to be the lowest level in
$\Lambda(<k) \sqcup \Lambda(k)$, then for the case $\Lambda(<k)=
\emptyset$, $s_0=k$ which means that $p \geq 1$ as in Section
(\ref{greaterk}). The results Sections (\ref{greaterk}),(\ref{lessk})
and (\ref{mixedk}) can then  be  summarised into

\begin{lemma} \label{three} 
Let $\cA =\Lambda(k) \sqcup \Lambda(>k)\sqcup \Lambda(<k)$ be an
inextendible antichain in the powerset lattice $\cB$. Let  $s_0$ be the
lowest level in $\Lambda(<k)\sqcup \Lambda(k)$, and let there exist at
least $k-s_0+1$ fine grained histories not contained in any of the
elements of $\Lambda(>k)\sqcup \Lambda(<k)$. Then, if every element of
$\cA$ has zero measure, so does $\Omega$. Hence $\cA$ is a quantum
cover of $\Omega$. 
\end{lemma} 

Lemma (\ref{three}), while quite general, clearly does not cover all
cases; while the conditions on $p$ are sufficient, they are
not necessary for proving $|\Omega|=0$. Relaxing the conditions on $p$
means that we can no longer use the equality of the measures
$|A_{\alpha_i}|$ and $|A_{\alpha_i} \sqcup A_{\alpha_j}|$'s in a
general way, independent of the details of $\cA$. A more general
technique than what we have used may exist, but eludes us at the
present. Instead, we deal with specific examples and show in every
case that the inextendible antichains we construct are quantum
covers\footnote{Indeed, in the tens of examples we have examined, we
  have not found a single  counterexample.}. While in
no way exhaustive, the following is a list of such examples.

\begin{enumerate}  
\item $\cA_1 \in \Big\{ \Lambda(k) \sqcup \Lambda(>k)\Big\}$,
$p=0$, $n>3$. \label{pzero}

Let $k=n-2$ and $p=0$. Consider the maximal antichain 
\begin{eqnarray} 
\mathcal{A}_1 &= & \{(n-1)\textrm{-level }: \{\bar A_1\},\{\bar A_2\};
\nn \\ 
&& \textrm{all} \,\, (n-2)\textrm{-level elements} \not\subset S_{n-2}(\bar
A_1)\bigcup S_{n-2}(\bar A_2)\},
\end{eqnarray} 
where $\{\bar A_\alpha\}$ denotes the set of all but the fine grained
history $A_\alpha$, $\alpha=1,2$. $|\{\bar A_\alpha\}|=0$ implies that
$|\Omega|=|A_1|=|A_2|$. Using the fact that  for any $a \in
S_{n-2}(\bar A_\alpha )$,  $|a|= |\bar A_\alpha \backslash a|$, and adding up
all the $(n-2)$-level elements as before gives 
\begin{equation} 
\frac{(4-n)(n-1)(n-2)}2\sum_{i=1}^n
|A_i|+\frac{(n-2)(n-3)}4\sum_{i,j=1}^n|A_i\sqcup A_j| = 
 2\sum_{i=3}^n|A_i|+|A_1|. 
 \end{equation} 
Thus, 
\beq|\Omega|=\frac{-2n+8}{(n-2)(n-3)}\sum_{i=3}^n|A_i|+\frac{-2n+5}{(n-2)(n-3)}|A_1|\leq
0\eeq

\item $\mathcal{A}_2 \in \Big\{
  \Lambda(k)\sqcup\Lambda(<k)\Big\},\quad p=1<k-s_0+1$, $n$ odd.   

The maximal antichain is defined as 
\bea 
\mathcal{A}_2 &=&\Big\{\frac{n+1}2-\textrm{level}: \{A_1,A_2,\cdots,
A_{\frac{n+1}2}\}, \{A_1,A_{\frac{n+3}2},\cdots A_n \}; \nonumber\\ &
& 2-\textrm{level}: \{A_i,A_j\}\quad i\in[2,\frac{n+1}2] \textrm{ and
} j\in[\frac{n+3}2,n] \Big\}.  
\eea 
Here, $k=\frac{n+1}2$, $s_0=2$, so
that $r=\frac{n-1}2$ which implies that $p=1 <r $. However, $\cA_2$
can also be viewed as an antichain in the set 
$\Big\{ \Lambda(k)\sqcup\Lambda(>k)\Big\}$ with $k=2$ and $p=0$. 

The $2$-level elements in $\cA_2$ imply that the $|A_i|$ are all equal for
$i\neq 1$.  Using $|A_2\sqcup\cdots \sqcup
A_{\frac{n+1}2}|=|A_1|=|A_{\frac{n+3}2}\sqcup\cdots \sqcup A_n|$, we
see that 
\begin{equation} 
\sum_{i_1,i_2\in[2,\frac{n+1}2]}|A_{i_1}\sqcup
A_{i_2}|=\sum_{j_1,j_2\in[\frac{n+3}2,n]}|A_{j_1}\sqcup
A_{j_2}|=|A_1|+\frac{(n-1)(n-5)}4|A_i|
\end{equation} 
Moreover,  $|A_1\sqcup A_i|=|A_1\sqcup A_j|$ for all
$i\in[2,\frac{n+1}2]$ and $j\in[\frac{n+3}2,n]$. Finally,  using the
$(\frac{n+1}2)$-level zero sets and Eqn \ref{identity}, 
\beq |A_1\sqcup A_i|=\frac{n-5}{n-1}|A_1|+|A_i|, \eeq 
which gives 
\beq 
|\Omega|=-|A_1|-\frac{(n-1)^2}2|A_i|\leq 0. 
\eeq

\item $\mathcal{A}_3 \in \Big\{
  \Lambda(k)\sqcup\Lambda(<k)\Big\},\quad p=1<k-s_0+1$, $m, l
  \equiv \frac{n-1}m$ positive integers.

This generalizes the previous example. Consider the maximal antichain
\begin{eqnarray} 
\mathcal{A}_3 &=&\Big\{(l+1)-\textrm{level}: \{A_1,A_2,\cdots,
A_{l+1}\}, \{A_1,A_{l+2},\cdots A_{2l+1}\}\cdots, \nonumber\\&
&\{A_1,A_{(m-1)l+2},\cdots A_{ml+1=n}\}; \nonumber\\ & &
2-\textrm{level}: \textrm{all} \{A_i,A_j\} \not \subset S_2(\textrm{(l+1)-level elements}), \Big\}
\end{eqnarray} 
The previous example is the special case, $m=2$. Note that the
$(l+1)$-level elements only overlap pairwise at $A_1$.  This
antichain  belongs to the $\Lambda(k)\sqcup\Lambda(<k)$ case, $p=1<r$
with $k=\frac{n-1}m+1$, and $s_0=2$ so that $ 
r=\frac{n-1}m$ and $p=1$. However, it can also be thought of as an element of
$\Lambda(k)\sqcup\Lambda(>k)$ with $k=2$ and $p=0$.

Retracing the arguments in the previous example, we see that 
all the $|A_i|$'s are equal for  $i\neq j$. Moreover,  
\begin{eqnarray} 
\sum_{i_1,i_2\in[(k-1)l+2,kl+1]}|A_{i_1}\sqcup A_{i_2}|&=&
\sum_{j_1,j_2\in[(k'-1)l+2,k'l+1]}|A_{j_1}\sqcup
A_{j_2}|\quad\forall k,k' \nonumber\\
&=&|A_1|+\frac{(n-1)(n-2m-1)}{m^2}|A_i|
\end{eqnarray}  
and  
\beq|A_1\sqcup A_i|=\left(1-\frac{2m}{n-1}\right)|A_1|+|A_i|. \eeq 
This implies that 
 \beq|\Omega|=(1-m)|A_1|+(n-1)(l+1-n)|A_i|\leq0 . 
\eeq
 
\item $\mathcal{A}_4 \in \Big\{
  \Lambda(k)\sqcup\Lambda(<k)\sqcup\Lambda(>k)\Big\}, \quad
  p=1<k-s_0+1$.
 
We consider the maximal antichain:
\begin{eqnarray} 
\mathcal{A}_4 &=&\Big\{(n-2)-\textrm{level}: \{\bar A_1,\bar 
A_2\}, \{\bar A_2, \bar A_3\}; \nonumber
\\& & l-\textrm{level}: \{A_1,A_2,A_{\alpha_1},\cdots, A_{\alpha_{l-2}}\}, \{A_2,A_3, A_{\alpha_1}\cdots, A_{\alpha_{l-2}} \},\nonumber\\& & \{A_2,\cdots,
A_{\alpha_{l-1}}\}, \, \, \alpha_i\in[4,\cdots,n];\nonumber\\
& & 2-\textrm{level}: \{A_1,A_3\} \Big\} 
\end{eqnarray} 
This antichain belongs to the mixed case with $p=1<r$ for $k=l$, and
since $s_0-2$,  $r=l-1$. Again, it can be viewed as an element of
$\Lambda(k)\sqcup\Lambda(<k)$ with $k=n-2$  and  $p=0$. 
 
From the $l$-level elements  we conclude that the $|A_i|$'s are all
equal for $i \neq 2$. Moreover, $|A_{\alpha_i}\sqcup
A_{\alpha_j}|=|A_1\sqcup A_{\alpha_k}|=|A_3\sqcup A_{\alpha_k}|$ for
all $\alpha_{i,j,k} \in[4,\cdots,n]$, and $|A_2\sqcup A_i|$'s are all
equal for all $i \neq 2$. $|A_1\sqcup A_2\sqcup A_3| = |A_2|$ gives us
the useful equality $|A_1|+|A_2|=|A_1\sqcup A_2|$, while $|A_1\sqcup
A_3\sqcup A_i|=|A_i|$ gives $|A_1|+|A_i|=|A_1\sqcup A_i|$. Using this
in the expression for a $l$-level zero set yields the equation
$(l-1)|A_1|+|A_2|=0$ which implies that $|A_1|=|A_2|=0$, and hence
$|\Omega|=0$.  Note that while we used the $l$-level and the $2$-level
elements we didn't need the $(n-2)$-level elements to prove this
result.
 
\end{enumerate}  

We thus obtain our main result, Lemma (\ref{result}), where $\cC$
includes all classes and specific examples of antichains considered so
far. Tens of other examples have also been examined, in every case,
verifying Conjecture \ref{conjecture}.

We have in this work used a particular characterisation of the
classical measure derived from zero sets to define our quantum
cover. In this we were motivated by the preclusion-based approaches to
quantum measure theory. Instead, one might want to consider the
``complementary'' property satisfied by a classical measure. Namely,
that given a cover $\{ O_i\}$ of a classical measure space $\Omega $,
$\sum_i |O_i|_c \geq |\Omega|_c$.  Indeed, as we now show,  this
property is satisfied by the $k$-level quantum cover.   

Eqn (\ref{klevelsum}) gives  the sum of  the measures of all $k$-level
elements and can be rearranged to  
\begin{equation} 
\frac{(n-2)!}{(k-2)!(n-k)!}\left(\frac{(2-k)(n-1)}{(k-1)}\sum_{i=1}^n
|A_i|+\frac{1}{2}\sum_{i,j=1}^n|A_i\sqcup A_j|\right). 
\end{equation}  
Using 
\beq\frac{(2-k)(n-1)}{(k-1)}=(2-n)+\frac{n-k}{k-1}\eeq we 
conclude that 
\begin{equation} 
\binom{n-2}{k-2}\biggl(|\Omega|+ \frac{n-k}{k-1}\sum_{i=1}^n 
|A_i|+\frac{1}{2}\sum_{i,j=1}^n|A_i\sqcup A_j|\biggr) \geq |\Omega|. 
\end{equation} 
Whether this result holds for all inextendible antichains or not is
still an open question. 


\section{The Peres-Kochen-Specker Theorem and Quantum Covers} 

In \cite{dgt}, Dowker and Ghazi-Tabatabai recast the work of Peres
\cite{pks} on the Kochen-Specker(KS) theorem \cite{ks} in the
framework of quantum measure theory and showed that it is consistent
with the anhomomorphic logic interpretation of quantum theory. We show
how their construction can help reinterpret the Peres-Kochen-Specker(PKS)
result in the language of covers.

We very briefly review the PKS set up as decribed in \cite{dgt} and
refer the reader to the orginal papers for more detail. One starts off
with the Peres Set(PS) of 33 rays in ${\mathbb R}^3$, which cluster
into 16 orthogonal bases, some overlapping with each other. Using
color labels for truth values, green is assigned for true and red for
false. An outcome $\gamma$ is a particular assignment of red or green
to each of the 33 rays in PS, and the base set $\Omega$ is the set of
all possible outcomes. In the classical realist picture, a particle
cannot simultaneously be in two orthogonal spin states. Thus, a
classical-realist path for the particle corresponds to a
``consistent'' coloring of the 33 rays in the PS; a simultaneous
assignment of green to just one out of three rays in every one of the
16 bases, and such that no pair of mutually orthogonal rays in the PS
are both green.  The PKS proof against classical realism is the proof
that there exists no ``consistent'' $\gamma$.

The PKS result is regarded as a definitive proof against realism, but
is more accurately a proof against classical realism. As shown in
\cite{dgt}, this subtle difference allows the anhomomorphic picture of
``quantum realism'' to be accomodated within the strictures of the
theorem.

The Dowker-Ghazi-Tabatabai version of the PKS result 
states:  
\begin{lemma}
{\bf (DGT):} Let $|.|$ be a measure on the space $\Omega$ of
colourings of PS that is zero valued on the PKS sets. Then there is no
preclusive classical coevent for this system.  
\end{lemma} 
The PKS sets correspond to the basis-wise and pair-wise sets of
inconsistent colorings: (a) if $B$ is a basis in the PS, then the set
$R_B \subset \Omega$ is the assignement of red to all three rays in
$B$ and (b) if $P$ is a mutually orthogonal pair of rays in the PS,
the set $G_P\subset \Omega$ is an assignment of green to both rays in
$P$.  The PKS collection is the set of all such subsets $R_B$ and
$G_P$ of $\Omega$. Classical realism then requires that each PKS set
is of measure zero, but the converse is not true, as implied by the
PKS theorem.  Note that the classical coevent referred to in the DGT
Lemma corresponds to a coevent with support on a single ``classical''
(or fine grained) element $\gamma$ of $\Omega$, which of course
exists, even though a classical-realistic $\gamma$ (corresponding to a
consistent colouring) does not.

In general, the non-existence of preclusive classical covents in the
anhomomorphic set up means that every $\gamma \in \Omega$ is contained
in a set of zero measure. Thus, there exists a collection of sets of
zero measure, which covers $\Omega$, but since the (normalised)
measure of $\Omega$ is $1$, it is not a quantum cover. Conversely, if
there exists a non-quantum covering of $\Omega$ of zero measure sets,
then there can be no preclusive classical coevent. That the PKS
collection is a non-quantum cover is obvious from the PKS theorem: the
statement that there exists a $\gamma \in \Omega$ such that $\gamma
\nin A$ for every $A$ in the PKS collection is equivalent to the
statement that there is a consistent coloring. Thus a quantum cover
avatar of the PKS-DGT result is: 
\begin{lemma} 
The PKS sets provide a non-quantum covering of $\Omega$. 
\end{lemma} 
While ruling out classical coevents does not come into conflict with the
anhomomorphic logic interpretation, one must also check if too
little information remains. Namely, is there only one (trivial)
primitive preclusive coevent whose support is $\Omega$? In the PKS set
up, it was shown in \cite{dgt} that this is not the case. More
generally, we note that non-triviality is an obvious consequence of
Lemma \ref{lemone}, for $\Omega$ of finite cardinality. Since
$|\Omega| =1$, all the elements in the $n-1$ level antichain cannot
have zero measure. Hence there exists at least one element $a$ in this
level, such that $|a| \neq 0$, and since it is not itself contained in
a zero measure set, $a \subseteq \mathrm{supp}(\Phi)$, where $\Phi$ is
a primitive preclusive coevent. Thus, $\mathrm{supp}(\Phi) \subset
\Omega$ is a strict inclusion, which means that  the statement
that ``something happens'', i.e. $|\Omega|=1$, can be refined to one with more
content. Therefore, in general, 
\begin{lemma} 
For any quantum system with an  $\Omega$ of finite cardinality, the
set of primitive preclusive coevents is non-trivial.  
\end{lemma}

Finally, as further support for our Conjecture \ref{conjecture} within
the PKS set up, we show that:
\begin{lemma} 
The  PKS collection $\cP$ does not form an inextendible antichain
 in the associated Boolean lattice $\cB$.
\end{lemma} 
\bproof Consider  the colorings of PS
\begin{eqnarray} 
\gamma&=& \{ \{u_1,u_2,u_3 \}\rightarrow red, \,\, \{u_4, \ldots u_{33}
\}\rightarrow  green \} \nn \\
\tgamma&=&  \{ \{u_1,u_2,u_3 \}\rightarrow green, \, \, \{u_4, \ldots u_{33}
\}\rightarrow  red \}, 
\end{eqnarray} 
where $B=\{ u_1,u_2,u_3\}$ is the basis $(001,010,100)$ of
\cite{pks,dgt}. Let $B^c=\{u_4, \ldots u_{33} \}$ which is the
remaining set of rays. Then, $\gamma \subset R_{B}$ and $\gamma
\subset G_{P_s}$, for all mutually orthogonal pairs $P_s$ in
$B^c$. If $B_i$, $i=1, \ldots 6$ are the 6 of the 16 bases
contained wholly within $B^c$, and $\tP_j$, $j=1,2,3$, the 3
mutually orthogonal pairs in $B$, then $\tgamma \subset
R_{B_i}$ for all $i$ and $\tgamma \subset G_{\tP_j}$ for all
$j$. Neither $\gamma$ nor $\tgamma$ are contained in any other sets in
$\cP$ besides. In particular, $\gamma \not\subset R_{B_i}$ for
all $i$ and $\gamma \not \subset G_{\tP_j}$ for all $j$ and
$\tgamma \not \subset R_{B}$ and $\tgamma \not \subset G_{P_s}$ for
all $s$.  Thus the set $\gamma\sqcup \tgamma \nsubseteq A$ for any $A
\in \cP$. Since the PKS sets have cardinality $> 2$,  $\gamma\sqcup
\tgamma$ cannot contain any set in $\cP$ either, and hence $\cP$
is not an inextendible antichain in $\cB$.  \eproof

\noindent{{\bf Acknowledgements:}} We would like to thank Fay Dowker,
Yousef Ghazi-Tabatabai and Rafael Sorkin for discussions on the PKS
theorem. This work was supported in part by the Royal Society-British
Council International Joint Project 2006/R2.

\section*{Appendix}

\subsection*{Proof of the Identity (\ref{identity})}

For convenience, we use the notation $[A_1](n)= \sum_{i=1}^n |A_i|$
and $[A_1A_2](n)= \sum_{i,j=1, i<j}^n |A_i\sqcup A_j|$, $[A_1A_2A_3](n)=
\sum_{i,j,k=1, i<j<k}^n |A_i\sqcup A_j \sqcup A_k|$, etc.  
The fact that $I_n(A_1, A_2 \ldots A_n)=0$ for $n>2$ means that 
\begin{eqnarray} 
|A_1\sqcup  \ldots \sqcup A_{n}| &=& 
\nn \\ 
&&  (-1)^0 [ A_1\ldots A_{n-1}](n) + (-1)^1[A_1 \ldots A_{n-2}](n) + \ldots
\nn \\ 
&& + (-1)^{n-3}[A_1A_2](n) + (-1)^{n-2}[A_1](n).   
\end{eqnarray}
For $n=3$, 
\begin{equation}
 |A_1\sqcup  A_2  \sqcup A_3|= |A_1 \sqcup A_2| + |A_2 \sqcup A_3| +
  |A_1 \sqcup A_3| - |A_1| -|A_2| -|A_3|, 
\end{equation} 
thus satisfying the identity (\ref{identity}). 
Now assume 
\begin{equation}\label{kexp} 
|A_1 \sqcup A_2 \ldots \sqcup A_k|=(2-k)[A](k)+ [AA'](k), \,  \forall
 \, 1 < k \leq n-1.    
\end{equation} 
Then, 
\begin{equation} 
|A_1\sqcup  \ldots \sqcup A_{n}| = Q \times [A_1](n) + P \times [A_1A_2](n) 
\end{equation} 
from symmetry. Consider the contribution to  $P$ from the term
$[A_1..A_k](n)$. Using (\ref{kexp}) we see that each term $|A_i\sqcup
A_j|$ for a given $i,j \in [1, \ldots n]$, $i\neq j$, 
appears $\binom{n-2}{k-2}$ times. Adding up the contributions from all
$k$ we get 
\begin{equation}
P = \sum_{l=1}^{n-2} (-1)^{l-1} \binom{n-2}{l} =1,   
\end{equation}   
where we have used the identity 
\begin{equation} \label{binomone} 
\sum_{l=0}^{n-2} (-1)^{l} \binom{n-2}{l}=0. 
\end{equation} 
Similarly, each term $|A_i|$ for a given $i\in [1, \ldots n]$ appears
in $[A_1..A_k](n)$ $\binom{n-1}{k-1}$ times. Thus, 
\begin{eqnarray} 
Q & = & (-1)^0 (2-n+1) \binom{n-1}{n-2} + (-1)^1 (2-n+2)
\binom{n-1}{n-3} \nn \\ 
&& \ldots (-1)^{n-k-1} (2-k) \binom{n-1}{k-1} + \ldots (-1)^{n-4}
(2-3) \binom{n-1}{2} + (-1)^{n-2}\nn  \\
&=& Q_1 +Q_2+ (-1)^{n-2},    
\end{eqnarray} 
where 
\begin{eqnarray}
Q_1&=&  - (2-n) \sum_{l=1}^{n-3}(-1)^l\binom{n-1}{l} \nn \\
Q_2 &=& - \sum_{l=1}^{n-3} (-1)^l  \binom{n-1}{l} \times  l . 
\end{eqnarray} 
Using (\ref{binomone}) and 
\begin{equation} 
\sum_{l=0}^{n-1} (-1)^l  \binom{n-1}{l} \times l=0, 
\end{equation} 
$Q$ simplifies to $(2-n)$, thus proving inductively, the identity
(\ref{identity}).  \hfill \eproof.

\subsection*{Proof of the identities (\ref{first}) and (\ref{second}). }
The identities (\ref{first}) and (\ref{second}) satisfied by the quantum
measure can be obtained from a strongly positive decoherence
functional, with $D(A,B)$, where $|A|=D(A,A)$ is the quantum measure.   
A decoherence functional is required to satisfy the following conditions  
\begin{eqnarray} 
D(A\sqcup B, C) &= & D(A,C)+D(B,C) \qquad  (\mathrm{Biadditivity})  \\ 
D(A,B) & = &  D^*(B,A) \qquad  (\mathrm{Hermiticity}) \\ 
D(A,A) & \geq &  0 \qquad  (\mathrm{Positivity})\label{positivity}  \\ 
D(\Omega,\Omega)&= & 1  \qquad (\mathrm{Normalisability}). 
\end{eqnarray} 
While Eqn (\ref{positivity}) is sufficient for satisfying the
probability interpretation for a classical partition, standard unitary
quantum mechanics satisfies the stronger condition of {\sl strong
positivity}, which allows a Hilbert space to be associated to $\Omega$
\cite{qrw}. Namely, for any $A \subset \Omega$, and $A=\sum_i
\gamma_i$ the matrix 
\begin{equation}
M_{ij} = \sum_{i,j} D(\gamma_i, \gamma_j) 
\end{equation} 
is positive, i.e. its eigenvalues are $\geq 0$. For two disjoint sets
$A$ and $B$, this gives us the important inequality,
\begin{equation}\label{dr}
D(A,A)D(B,B) \geq |D(A,B)|^2 \Rightarrow  D(A,A)D(B,B) \geq D_R(A,B)^2,    
\end{equation} 
where $D_R(A,B)= {\rm{Re}}D(A,B)$. Now, biadditivity tells us that  
\begin{equation}
D(A\sqcup B, A\sqcup B) =  D(A,A)+ D(B,B) + 2 D_R(A,B).   
\end{equation}  
Combining this with (\ref{dr}), 
\begin{equation}\label{ineq} 
(\sqrt{D(A,A)}- \sqrt{D(B,B)})^2 \leq D(A\sqcup B, A\sqcup B)    \leq
  (\sqrt{D(A,A)}+\sqrt{D(B,B)})^2.  
\end{equation} 
From (\ref{dr}) 
\begin{equation} 
D(A,A)=0 \Rightarrow |D(A,B)|=0 \Rightarrow D(A\sqcup B, A\sqcup B)=
D(B,B),    
\end{equation}
thus proving (\ref{second}). Next, from (\ref{ineq})    
\begin{equation}
 D(A\sqcup B, A\sqcup B)=0 \Rightarrow (\sqrt{D(A,A)}-
 \sqrt{D(B,B)})^2=0  \Rightarrow D(A,A)=D(B,B),  
\end{equation} 
thus proving (\ref{first}).

\end{document}